\documentclass[aps,pra,reprint,superscriptaddress,amsmath,amssymb,showpacs]{revtex4-1}
\usepackage{graphicx}

\begin{document}


\title{Optimality for indecomposable entanglement witnesses}

\author{Kil-Chan Ha}
\affiliation{Faculty of Mathematics and Statistics, Sejong University, Seoul 143-747, Korea}

\author{Seung-Hyeok Kye}
\affiliation{Department of Mathematics and Institute of Mathematics, Seoul National University, Seoul 151-742, Korea}
\date{\today}

\begin{abstract}
We examine various notions related with the optimality for
entanglement witnesses arising from Choi type positive linear maps.
We found examples of optimal entanglement witnesses which are
non-decomposable, but which are not \lq non-decomposable optimal
entanglement witnesses\rq\ in the sense of [M. Lewenstein, B. Kraus,
J. Cirac, and P. Horodecki, Phys. Rev. A {\bf 62}, 052310 (2000)].
We suggest to use the term {\sl PPTES witness} and {\sl optimal
PPTES witness} in the places of \lq non-decomposable entanglement
witness\rq\ and \lq non-decomposable optimal entanglement
witnesses\rq\ in order to avoid possible confusion. We also found
examples of non-extremal optimal entanglement witnesses which are indecomposable.
\end{abstract}


\maketitle


\section{Introduction}

Quantum entanglement is now considered as the main key resource for applications to
quantum information and quantum computation theory. 
One of the major research topics in the theory of entanglement is,
of course, how to distinguish entanglement from separable states.
For this purpose, positive linear maps are known to be the most
complete tools \cite{horo-1} among various criteria. This criterion
for separability using positive maps is equivalent to the duality
theory \cite{eom-kye} between positivity of linear maps and
separability of block matrices, through the Jamio\l kowski-Choi
isomorphism \cite{choi75-10,jami}.
In this sense, we need a positive linear map to
detect entanglement. This is formulated as the notion of
entanglement witness \cite{terhal} that is just a positive linear
map which is not completely positive, under the isomorphism.
We refer to \cite{ssz,ZB} for systematic approaches
to the duality using the Jamio\l kowski-Choi isomorphism.

An entanglement witness which detects a maximal set of entanglement
is said to be optimal, as was introduced in \cite{lew00}. The notion
of optimality may be explained in terms of facial structures of the
convex cone $\mathbb P_1$ consisting of all positive linear maps
between matrix algebras. In fact, it was shown \cite{kye_dec_wit}
that a positive map $\phi$ is an optimal entanglement witness if and
only if the smallest face of $\mathbb P_1$ containing $\phi$ has no
completely positive linear map. See also Ref.~\cite{sar}. Therefore,
the most natural candidates of optimal entanglement witnesses are
extremal positive maps which are not completely positive. In spite
of its importance, the facial structure of the cone $\mathbb P_1$
is very far from being understood even in the low dimensional cases.
When both the domain and the range are the $2\times 2$ matrix
algebra, all extreme points of the convex set consisting of unital
positive maps had been found in the sixties \cite{stormer}. The
whole facial structures of this convex set is completely understood
by the second author \cite{kye-2by2_II}. See also Ref.~\cite{byeon-kye}.
Another sufficient condition for optimality is the notion of the
spanning property, as was introduced in \cite{lew00}. This is very
useful, because the spanning property is much easier to verify than
the optimality itself. It turns out \cite{kye_ritsu} that a positive
map $\phi$ has the spanning property if and only if the smallest
exposed face of the cone $\mathbb P_1$ containing $\phi$ has no
completely positive map.

Recall that a convex subset $F$ of a convex set $C$ is said to be a face if the following condition holds:
If a convex combination of two points $x,y\in C$ belongs to $F$ then $x$ and $y$ themselves belong to $F$.
A face  $F$ of $C$ is said to be an exposed face if it is the intersection of $C$ and a hyperplane.
We will see an example of a face which is not exposed through the discussion. See FIG. 1.

For the decomposable case, several necessary and/or sufficient
conditions for optimality are known, and there are progresses to
characterize optimal decomposable entanglement witnesses. See
Refs.~\cite{aug,asl,kye_dec_wit} for examples. In the case of indecomposable entanglement witnesses,
a condition for optimality has been found \cite{xia} recently, and examples of optimal entanglement witnesses
without the spanning property were given.
Nevertheless, we have still few kinds of examples for optimal
entanglement witnesses arising from indecomposable maps.
We note that the Choi type positive maps are one of the main resources for indecomposable positive maps.
The primary purpose of this note is to analyze those maps between $3\times 3$ matrix algebras,
and examine the relations between extremeness, spanning property and optimality.

We note that a positive map $\phi$ detects entanglement with positive partial transposes if and only if it is indecomposable.
An indecomposable positive map $\phi$ is said to be a non-decomposable optimal entanglement witness (nd-OEW) in \cite{lew00}
if it detects a maximal set of PPTES.
But, it is not clear at all that an optimal entanglement witness which is non-decomposable is really nd-OEW in the sense of \cite{lew00}. We found that this is not the case.
In order to avoid such confusion, we use the following terminology in this note. A positive linear map $\phi$ is said to
\begin{itemize}
\item
be {\sl co-optimal} if the smallest face of $\mathbb P_1$ containing $\phi$ has no completely copositive map.
\item
be {\sl bi-optimal} if it is optimal and co-optimal.
\item
have the {\sl co-spanning property} if the smallest exposed face of $\mathbb P_1$ containing $\phi$ has no completely copositive map.
\item
have the {\sl bi-spanning property} if it has both the spanning and co-spanning property.
\end{itemize}
It is clear that $\phi$ is co-optimal (respectively has the
co-spanning property) if and only if the composition
$\phi\circ{\text{\rm t}}$ with the transpose map ${\text{\rm t}}$ is
optimal (respectively has the spanning property). If we use the
Jamio\l kowski-Choi isomorphism, then a self-adjoint block matrix
$W$ is co-optimal (respectively has the co-spanning property) if and
only if the partial transpose $W^\Gamma$ is optimal (respectively
has the spanning property). It is also clear that $\phi$ is
bi-optimal (respectively has the bi-spanning property) if and only
if the smallest face (respectively the smallest exposed face) of
$\mathbb P_1$ containing $\phi$ has no decomposable map. Therefore,
$\phi$ is an nd-OEW in the sense of \cite{lew00} if and only if it
is bi-optimal. We note that if $\phi$ is bi-optimal then it is
automatically indecomposable. We will present examples of
indecomposable optimal positive linear maps which are not
bi-optimal. Since an optimal decomposable entanglement witness is
completely copositive, it is never co-optimal. Therefore, the notions of co-optimality and co-spanning
are useful only for indecomposable entanglement witnesses.


For nonnegative real numbers $a,b$ and $c$, the Choi type map is given by
\begin{widetext}
\[
\Phi[a,b,c](X)=\\
\begin{pmatrix}
ax_{11}+bx_{22}+cx_{33} & -x_{12} & -x_{13} \\
-x_{21} & cx_{11}+ax_{22}+bx_{33} & -x_{23} \\
-x_{31} & -x_{32} & bx_{11}+cx_{22}+ax_{33}
\end{pmatrix},
\]
\end{widetext}
for $X=[x_{ij}]\in M_3$, where $M_3$ denotes the $C^*$-algebra of
all $3\times 3$ matrices over the complex field $\mathbb C$. Choi \cite{choi72}
showed that the map $\Phi[1,2,2]$ is a $2$-positive linear map which is not completely positive.
This is the first known example to distinguish $n$-positivities for different $n=2,3,\dots$.
The map $\Phi[1,0,\mu]$ with
$\mu\ge 1$ is also the first example of an indecomposable positive linear map \cite{choi75} in the literature,
and the map $\Phi[1,0,1]$ is extremal \cite{choi-lam}, that is, generates an extremal ray of the cone $\mathbb P_1$.
Later, it was shown \cite{tomiyama} that this map $\Phi[1,0,1]$
is not the sum of a $2$-positive map and a $2$-copositive map. See also Ref.~\cite{ha-atomic}.
The map $\Phi[1,0,1]$ is usually called the Choi map.
The maps $\Phi[a,b,c]$ have been considered in \cite{cho-kye-lee} to distinguish various notions of positivity.
See also \cite{xia, ando,ber,ck-Choi,ck--1,ck,osaka_93,arvind,singh,tomiyama,yamagami,cw} for another variations of the Choi map.
It is known \cite{cho-kye-lee} that the map $\Phi[a,b,c]$ is positive if and only if the condition
\begin{equation}\label{pos}
a+b+c\ge 2,\qquad 0\le a\le 1\Longrightarrow bc\ge (1-a)^2
\end{equation}
holds.
Note that $\Phi[1,0,1]$ is optimal by the extremeness. It is also well known that $\Phi[1,0,1]$ has not the spanning property,
as was observed in \cite{kye-canad}. See also Ref.~\cite{korb}. It is also known to
have the co-spanning property \cite{choi_kye}.
Recently, the authors \cite{ha+kye_indec-witness} have shown that
if $0<a< 1$ and the equalities hold in the both inequalities in (\ref{pos}) then $\Phi[a,b,c]$ has the bi-spanning property.
We note that the Choi matrix $C_\Phi=\sum_{i,j=0}^2 |i\rangle \langle j|\otimes \Phi(|i\rangle \langle j|)$ of the map $\Phi[a,b,c]$ is given by
\begin{equation}\label{choi_matrix}
W[a,b,c]=\left(
\begin{array}{ccccccccccc}
a     &\cdot   &\cdot  &\cdot  &-1     &\cdot   &\cdot   &\cdot  &-1     \\
\cdot   &c &\cdot    &\cdot    &\cdot   &\cdot &\cdot &\cdot     &\cdot   \\
\cdot  &\cdot    &b &\cdot &\cdot  &\cdot    &\cdot    &\cdot &\cdot  \\
\cdot  &\cdot    &\cdot &b &\cdot  &\cdot    &\cdot    &\cdot &\cdot  \\
-1     &\cdot   &\cdot  &\cdot  &a     &\cdot   &\cdot   &\cdot  &-1     \\
\cdot   &\cdot &\cdot    &\cdot    &\cdot   &c &\cdot &\cdot    &\cdot   \\
\cdot   &\cdot &\cdot    &\cdot    &\cdot   &\cdot &c &\cdot    &\cdot   \\
\cdot  &\cdot    &\cdot &\cdot &\cdot  &\cdot    &\cdot    &b &\cdot  \\
-1    &\cdot   &\cdot  &\cdot  &-1     &\cdot   &\cdot   &\cdot  &a
\end{array}
\right).
\end{equation}

In the next section, we examine the above mentioned properties for boundary points of the convex body determined by the condition
(\ref{pos}), and discuss the result in the final section.

\section{Facial structures and Optimality}
Before going further, we note that the six properties, optimal, co-optimal, bi-optimal, spanning, co-spanning and bi-spanning
are properties depending on the faces:
If $\phi_1$ and $\phi_2$ determine the same smallest face containing them, then they are interior points of a common face,
and share the each property, because the properties are described in terms of faces.
Therefore, we can say that a face itself has one among six properties without confusion, and this means that every
interior point of the face satisfies the property.
It is also clear that a face has a property, then every subface also has the same property.
Hence, if a point $\phi$ does not have a property then every interior point in the face containing $\phi$ does not have the property.
Therefore, we need to clarify the facial structures of the $3$-dimensional convex body itself determined by
(\ref{pos}). It should be noted that the face of the convex body need not give rise to a real face of the convex cone
$\mathbb P_1$. Nevertheless, an interior point of a face of the convex body gives rise to an interior point of the face of the cone $\mathbb P_1$
determined by the corresponding map.

First of all, the convex body has the following four $2$-dimensional faces:
\begin{itemize}
\item
$f_{\rm ab}=\{(a,b,c): c=0,\ a+b\ge 2,\ a\ge 1\}$,
\item
$f_{\rm ac}=\{(a,b,c): b=0,\ a+c\ge 2,\ a\ge 1\}$,
\item
$f_{\rm bc}=\{(a,b,c): a=0,\ bc\ge 1\}$,
\item
$f_{\rm abc}=\{(a,b,c): a+b+c=2,\\\phantom{XXXXXX}  0\le a\le 1\Longrightarrow bc\ge (1-a)^2\}$.
\end{itemize}
We note \cite{cho-kye-lee} that $\Phi[a,b,c]$ is completely positive if and only if
$a\ge 2$, and it is completely copositive if and only if $bc\ge 1$. Therefore, the face $f_{\rm abc}$ has the completely positive map
$\Phi[2,0,0]$ and the completely copositive map $\Phi[0,1,1]$, and so $f_{\rm abc}$ is neither optimal nor co-optimal.
It is also easy to examine the optimality for the first three cases.
For example, if $a>2$ then the map $\Phi[a,0,0]$
is written by
$$
\Phi[a,0,0]=\Phi[2,0,0]+(a-2)D,
$$
where $D$ is the diagonal map which send $[x_{ij}]$ to the diagonal matrix with the diagonal entries $(x_{11},x_{22},x_{33})$. The map $D$ is
both completely positive and completely copositive. This means that the map $\Phi[a,0,0]$
never satisfy optimality and co-optimality. Therefore, every interior point in the $2$-dimensional
faces $f_{\rm ac}$ and $f_{\rm ab}$ never satisfy above properties. By the same argument, this is also the case for the face $f_{\rm bc}$.

We note that the convex body has also the following five $1$-dimensional faces which are on the $a$-axis, $ab$-plane or $ac$-plane:
\begin{itemize}
\item
$e_{\rm a}=\{(a,0,0): a\ge 2\}$,
\item
$e_{\rm b}=\{(1,b,0): b\ge 1\}$,
\item
$e_{\rm c}=\{(1,0,c): c\ge 1\}$,
\item
$e_{\rm ab}=\{(a,b,0): a+b=2,\, 1\le a \le 2\}$,
\item
$e_{\rm ac}=\{(a,0,c): a+c=2,\, 1\le a \le 2\}$.
\end{itemize}
Among them, we have already seen that the face $e_{\rm a}$ is neither optimal nor co-optimal.
This is also the case for $e_{\rm b}$ and $e_{\rm c}$, since it is possible to
subtract a map which is both completely positive and completely copositive.
It is also clear that neither  $e_{\rm ab}$ nor $e_{\rm ac}$ is optimal.
In order to find other $1$-dimensional faces, we note that the parametrization
$$
(a(t),\,b(t),\,c(t))=\frac1{1-t+t^2}((1-t)^2,\,t^2,\,1),\quad 0<t<\infty
$$
satisfies the condition
$$
a(t)+b(t)+c(t)=2,\ 0\le a(t)\le 1,\ b(t)c(t)=(1-a(t))^2,
$$
as was considered in \cite{ha+kye_indec-witness}. For each fixed positive number $t>0$ with $t\neq 1$, the line segment given by
\begin{itemize}
\item
$e_t=\left\{\left(1-s,\,st,\, s/ t\right):  t/{(t^2-t+1)}\le s \le 1\right\}$
\end{itemize}
lies on the surface $bc= (1-a)^2$ for $0\le a < 1$, and connects the point $(a(t),\,b(t),\,c(t))$ to the point $(0,t,1/t)$.
This gives us $1$-dimensional faces $e_t$ for each $t>0$ with  $t\neq 1$. Note that $\Phi[0,t,1/t]$ is
completely copositive  for each $t>0$, and so it is clear that $e_t$ is not co-optimal.

It remains to list up $0$-dimensional faces as follows:
\begin{itemize}
\item
$v_{(2,0,0)},\ v_{(1,0,1)},\ v_{(1,1,0)}$,
\item
$v_{(a(t),b(t),c(t))}$ for $t>0$ and $t\neq 1$,
\item
$v_{(0,t,1/t)}$ for $t>0$.
\end{itemize}

\begin{figure}[h!]
\begin{center}
\includegraphics[scale=0.5]{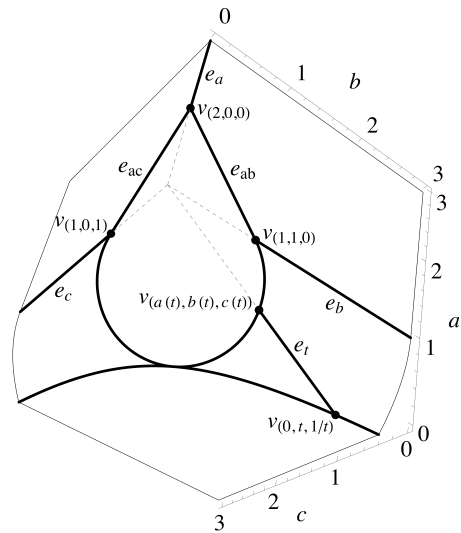}
\end{center}
\caption{Part of convex body determined by Eq.~\eqref{pos}.
The smallest face containing $v_{(1,1,0)}$ is itself, but the smallest exposed face containing it is $e_{\rm ab}$.
The straight lines containing the faces $e_a,\,e_b,\,e_c$, and $e_t$ meet each other at the point $(1,0,0)$
which is not in the convex body. }
\end{figure}
So far, we have seen that the faces $f_{\rm ab}$, $f_{\rm ac}$, $f_{\rm bc}$, $f_{\rm abc}$, $e_{\rm a}$, $e_{\rm b}$, and $e_{\rm c}$
are neither optimal nor co-optimal. Therefore, they have neither the spanning property nor co-spanning property.
We test the other faces. First of all, we show that $e_t$ and $v_{(0,t,1/t)}$ have the spanning properties.
To do this, it suffices to consider the case when $(a,b,c)$ satisfies the condition
\begin{equation}\label{cccc}
0\le a<1,\qquad bc=(1-a)^2,\qquad a+b+c>2.
\end{equation}

We recall \cite{kye_ritsu} (see also Ref.~\cite{lew00}) that $\phi\in\mathbb P_1$ has the spanning property if and only if the set
$$
P[\phi]:=\{\xi\otimes\eta\in \mathbb C^m\otimes \mathbb C^n:\langle \xi \otimes \eta|C_{\phi}|\xi \otimes \eta\rangle =0\}
$$
spans the whole space $\mathbb C^m\otimes\mathbb C^n$, where $C_{\phi}$ is the Choi matrix of $\phi$. We define vectors in $\mathbb C^3$ as follows
\begin{equation}\label{prod}
\begin{aligned}
|\xi_{\theta,\sigma}^0\rangle=&e^{i\theta}b^{1/4}|1\rangle+e^{i\sigma}c^{1/4}|2\rangle,\\
|\xi_{\theta,\sigma}^1\rangle=&e^{i\theta}b^{1/4}|2\rangle+e^{i\sigma}c^{1/4}|0\rangle,\\
|\xi_{\theta,\sigma}^2\rangle=&e^{i\theta}b^{1/4}|0\rangle+e^{i\sigma}c^{1/4}|1\rangle,\\
|\eta_{\theta,\sigma}^0\rangle=&e^{-i\theta}(bc)^{1/4}|1\rangle+e^{-i\sigma}b^{1/2}|2\rangle,\\
|\eta_{\theta,\sigma}^1\rangle=&e^{-i\theta}(bc)^{1/4}|2\rangle+e^{-i\sigma}b^{1/2}|0\rangle,\\
|\eta_{\theta,\sigma}^2\rangle=&e^{-i\theta}(bc)^{1/4}|0\rangle+e^{-i\sigma}b^{1/2}|1\rangle.\\
\end{aligned}
\end{equation}
Then, it is easy to check that
\[
\begin{aligned}
&\langle \xi_{\theta,\sigma}^k\otimes \eta_{\theta,\sigma}^k|C_{\Phi}|\xi_{\theta,\sigma}^k\otimes \eta_{\theta,\sigma}^k\rangle\\
=&\langle \xi_{\theta,\sigma}^k\otimes \eta_{\theta,\sigma}^k|W[a,b,c]|\xi_{\theta,\sigma}^k\otimes \eta_{\theta,\sigma}^k\rangle\\
=&-2(1-a)bc^{1/2}+2b^{3/2}c
\end{aligned}
\]
for all $k=1,2,3$, and $\langle \xi_{\theta,\sigma}^k\otimes \eta_{\theta,\sigma}^k|C_{\Phi}|\xi_{\theta,\sigma}^k\otimes \eta_{\theta,\sigma}^k\rangle=0$, whenever the condition~\eqref{cccc} holds. Therefore,  the vectors $|\xi_{\theta,\sigma}^k\otimes \eta_{\theta,\sigma}^k\rangle$ belong to $P[\Phi[a,b,c]]$ for all $k=1,2,3$
whenever the condition (\ref{cccc}) holds.
We take $\sigma_1=0,\, \sigma_2=\pi/2$ and
$\sigma_3=\pi$, and consider the $9\times 9$ matrix whose columns are
nine vectors $|\xi_{0,\sigma_{\ell}}^k\otimes \eta_{0,\sigma_{\ell}}^k\rangle$ for $k,\,\ell=1,2,3$. Then the
determinant of $M$ is given by
$$
|\det M|=128\,b^{\frac 92}c^{\frac 94}
$$
which is nonzero. This shows that $e_t$ and $v_{(0,t,1/t)}$ have the spanning properties.

Next, we consider the $0$-dimensional face $v_{(2,0,0)}$.  We see that the smallest
exposed face $F$ containing $v_{(1,0,1)}$ already contains $v_{(2,0,0)}$ in the FIG.\,1.(See Ref.~\cite{kye_ritsu} for more general approach). We have seen \cite{choi_kye}
that $\Phi[1,0,1]$ has the co-spanning property,
and so $F$ has no completely copositive map. This show that $v_{(2,0,0)}$ has the co-spanning property,
and so $e_{\rm ab}$ and $e_{\rm ac}$ also have the co-spanning properties.

We summarize the result as follows:
\begin{widetext}\begin{center}
\begin{table}[h!]
\begin{tabular}{ccccccccccc}
  \hline\hline
 & & &\multicolumn{3}{c}{(Co-)Spanning property} & & &\multicolumn{3}{c}{(Co-)Optimality}\\\cline{4-6}\cline{9-11}
Faces & &  &Span. &   Co-span. &Bi-span.& & &Opt. &Co-opt. &Bi-opt.\\\hline
$f_{\rm ab}, f_{\rm ac}, f_{\rm bc}, f_{\rm abc}, e_{\rm a}, e_{\rm b},e_{\rm c}$
& & &N&N&N& &  &N&N&N\\
 $e_{\rm ab},e_{\rm ac},v_{(2,0,0)}$
& & &N&Y&N& & &N&Y&N\\
 $e_t, v_{(0,t,1/t)}$
& & &Y&N&N& & &Y&N&N\\
$v_{(1,0,1)},v_{(1,1,0)}$
& & &N&Y&N& & &Y&Y&Y\\
$v_{(a(t),b(t),c(t))}$
& & &Y&Y&Y& & &Y&Y&Y\\
 \hline\hline
\end{tabular}\caption{Summary of (co-)optimality and (co-)spanning property for faces of the convex body illustrated in Fig. 1.}
\end{table}
\end{center}
\end{widetext}

We note \cite{cho-kye-lee} that the map $\Phi[a,b,c]$ is decomposable if and only if the following condition
$$
0\le a\le 2\Longrightarrow bc\ge \left(\frac{2-a}2 \right)^2
$$
holds. Therefore, we see that interior points of the following faces
$$
e_{\rm b},\,e_{\rm c},\,e_{\rm ab},\,e_{\rm ac},\,e_t,\,
v_{(1,0,1)},\,v_{(1,1,0)},\,v_{(a(t),b(t),c(t))}
$$
give rise to indecomposable positive maps. We note that every interior point of the face $e_t$ gives rise to
an example of an indecomposable optimal entanglement witness which is not bi-optimal. So, this is not \lq nd-OEW\rq\
in the sense of \cite{lew00}. If we consider the composition by the transpose map then the faces $e_{\rm ab}$ and $e_{\rm ac}$
play the exactly same role. They also provide us examples of non-extremal entanglement witnesses with the spanning property.
On the other hand, the Choi maps $v_{(1,0,1)}$ and $v_{(1,1,0)}$ are extremal entanglement witnesses without the spanning property.
Therefore, we see that two sufficient conditions, extremeness and spanning property, for the optimality is logically
independent.

\section{conclusions}

In this note, we considered Choi type positive maps between $3\times 3$ matrices, and determined
their optimality, co-optimality, spanning property and co-spanning property.
We have seen that even though a non-decomposable entanglement witness is optimal, it need not to be a \lq non-decomposable
optimal entanglement witness\rq\  in the sense of \cite{lew00}.
Because a positive map detects a PPTES if and only if it is indecomposable, we suggest to use
the term {\sl PPTES witness} in the place of non-decomposable entanglement witness, and use the term {\sl optimal
PPTES witness} in the place of nd-OEW. In other word, we say that a positive map is an optimal PPTES witness when it
is bi-optimal. This is very natural since a positive map detects a maximal set of PPTES if and only if it is bi-optimal.

Optimality is not so easy to determine for a given positive linear map, because we do not know the whole facial structures
of the convex cone $\mathbb P_1$ consisting of all positive maps. The spanning property is stronger than optimality and relatively easy to
check. Another sufficient condition for optimality is extremeness. We also showed that spanning property and extremeness are
logically independent.

\begin{acknowledgments}
This work was partially supported by the Basic Science Research Program through the
National Research Foundation of Korea(NRF) funded by the Ministry of Education, Science
and Technology (Grant No. NRFK 2011-0006561 to K. -C. Ha and Grant No. NRFK 2012-0000939 to S.-H. Kye)
\end{acknowledgments}


\begin{thebibliography}{99}

\bibitem{horo-1} M. Horodecki, P. Horodecki, and R. Horodecki,
Phys. Lett. A {\bf 223}, 1 (1996).

\bibitem{eom-kye} M.-H. Eom and S.-H. Kye,
Math. Scand. {\bf 86}, 130 (2000).

\bibitem{choi75-10} M.-D. Choi,
Linear Algebra Its Appl. {\bf 10}, 285 (1975).

\bibitem{jami} A. Jamio\l kowski,
Math. Phys. {\bf 5}, 415 (1974).

\bibitem{terhal} B. M. Terhal,
Phys. Lett. A {\bf 271}, 319 (2000).

\bibitem{ssz} \L. Skowronek, E. St\o rmer, and K. Zyczkowski,
Math. Phys. {\bf 50}, 062106 (2009).

\bibitem{ZB} K. \.Zyczkowski and I. Bengtsson,
Open Syst. Inf. Dyn. {\bf 11}, 3 (2004).

\bibitem{lew00} M. Lewenstein, B. Kraus, J. Cirac, and P. Horodecki,
Phys. Rev. A {\bf 62}, 052310 (2000).

\bibitem{kye_dec_wit} S.-H. Kye,
Rep. Math. Phys (to be published), e-print arXiv:1108.0456.

\bibitem{sar} G. Sarbicki,
e-print arXiv:0905.0778.

\bibitem{stormer} E. St\o rmer,
Acta Math. {\bf 110}, 233 (1963).

\bibitem{kye-2by2_II} S.-H. Kye,
Linear Algebra Its Appl. {\bf 362}, 57 (2003).

\bibitem{byeon-kye} E.-S. Byeon and S.-H. Kye,
Positivity {\bf 6}, 369 (2002).

\bibitem{kye_ritsu} S.-H. Kye,
e-print arXiv:1202.4255.

\bibitem{asl} R. Augusiak, G. Sarbicki and M. Lewenstein,
Phys. Rev. A {\bf 84}, 052323 (2011).

\bibitem{aug} R. Augusiak, J. Tura, and M. Lewenstein,
 J. Phys. A {\bf 44},  212001 (2011).

\bibitem{xia} X.-F. Qi and J.-C. Hou,
Phys. Rev. A {\bf 85}, 022334 (2012).

\bibitem{choi72} M.-D. Choi,
Canad. Math. J. {\bf 24}, 520 (1972).

\bibitem{choi75} M.-D. Choi,
Linear Algebra Its Appl. {\bf 12}, 95 (1975).

\bibitem{choi-lam}  M.-D. Choi and T.-T. Lam,
Math. Ann. {\bf 231}, 1 (1977).

\bibitem{tomiyama} K. Tanahashi and J. Tomiyama,
Canad. Math. Bull. {\bf 31}, 308 (1988).

\bibitem{ha-atomic} K.-C. Ha,
Publ. Res. Inst. Math. Sci. {\bf 34}, 591  (1998).

\bibitem{cho-kye-lee} S.-J. Cho, S.-H. Kye, and S. G. Lee,
Linear Algebra Its Appl. {\bf 171}, 213 (1992).


\bibitem{ando} T. Ando, Seminar note, 1985.

\bibitem{osaka_93} H. Osaka,
Linear Algebra Its Appl. {\bf 153}, 73 (1991).

\bibitem{yamagami} S. Yamagami,
Proc. Amer. Math. Soc {\bf 118}, 521 (1993).

\bibitem{ber} R. A. Bertlmann, K. Durstberger, B. C. Hiesmayr, and P. Krammer,
Phys. Rev. A {\bf 72},  052331 (2005).

\bibitem{ck-Choi} D. Chru\'{s}ci\'{n}ski and A. Kossakowski,
 J. Phys. A {\bf 41}, 145301 (2008).

\bibitem{ck--1} D. Chru\'{s}ci\'{n}ski and A. Kossakowski,
Commun. Math. Phys. {\bf 290}, 1051 (2009).

\bibitem{ck} D. Chru\'{s}ci\'{n}ski and A. Kossakowski,
Phys. Lett. A {\bf 373}, 2301 (2009).


\bibitem{arvind} R. Sengupta and Arvind,
e-print arXiv:1106.4279.

\bibitem{singh} A. I. Singh,
e-print arXiv:1201.0250.


\bibitem{cw} D. Chru\'{s}ci\'{n}ski and F. A. Wudarski,
e-print arXiv:1204.5283.


\bibitem{kye-canad} S.-H. Kye,
Canad. Math. Bull. {\bf 39}, 74 (1996).

\bibitem{korb} J. K. Korbicz, M. L. Almeida, J. Bae, M. Lewenstein, A. Acin,
Phys. Rev. A {\bf 78}, 062105 (2008).

\bibitem{choi_kye} H.-S. Choi and S.-H. Kye,
J. Korean Math. Soc. {\bf 49}, 623 (2012).

\bibitem{ha+kye_indec-witness} K.-C. Ha and S.-H. Kye,
 Phys. Rev. A {\bf 84}, 024302 (2011).
\end{thebibliography}

\end{document}